\documentstyle[12pt,epsfig]{article}

\newcommand{\beq}{\begin{equation} }
\newcommand{\eeq} {\end{equation} }

\begin{document}\begin{titlepage}

\begin{flushright}
%\preprint{
OSU-HEP-02-06\\
April 2002\\
%}
\end{flushright}

\vspace*{2.0cm}
\begin{center}
{\Large {\bf Finite Grand Unified Theories\\[-0.05in]
\,\,\,\,and The Quark Mixing Matrix} }

\vspace*{2cm}
 {\large {\bf K.S. Babu,\footnote{E-mail address: babu@okstate.edu}
 Ts. Enkhbat\footnote{E-mail address: enkhbat@okstate.edu}
 and I. Gogoladze\footnote{On a leave of absence from:
Andronikashvili Institute of Physics, GAS, 380077 Tbilisi,
Georgia.  \\ E-mail address: ilia@hep.phy.okstate.edu}}}

 \vspace*{1cm}
{\small \it Department of Physics, Oklahoma State University\\
Stillwater, OK~74078, USA }
\end{center}

 \vspace*{2.0cm}

\begin{abstract}

In $N=1$ super Yang--Mills theories, under certain conditions
satisfied by the spectrum and the Yukawa couplings, the beta
functions will vanish to all orders in perturbation theory.  We
address the generation of realistic quark mixing angles and masses
in such finite Grand Unified Theories. Working in the context of
finite SUSY $SU(5)$, we present several examples with realistic
quark mixing matrices. Non-Abelian discrete symmetries are found
to be important in satisfying the conditions for finiteness. Our
realistic examples are based on permutation symmetries and the
tetrahedral symmetry $A_4$.  These examples enable us to address
questions such as the decay rate of the proton in finite GUTs.

\end{abstract}
\end{titlepage}

\newpage

\section{Introduction}

The Standard Model (SM) has confronted the experimental data with
amazing success.  Nevertheless, it is still considered by many as
a low energy limit of a deeper underlying theory.  The gauge
hierarchy problem and the proliferation of parameters, especially
in the fermion masses and mixings, are two of the major reasons
for this belief.  Various extensions of the SM address different
aspects of these problems.  Typically these extensions involve
higher symmetries.  While the number of effective parameters in
these models with higher symmetries (e.g. grand unified theories)
might be smaller than the SM, because of the necessity to break
the higher symmetry, the actual number of parameters are often
larger than the SM.  Thus it becomes natural to ask whether it is
possible to have a theory in which there are fewer number of
parameters.

Indeed, there exists a certain class of supersymmetric Yang--Mills
theories, where one may achieve this goal. These are the
so--called finite theories wherein the $\beta$ functions for the
gauge coupling and the Yukawa couplings vanish to all orders in
perturbation theory.  Certain conditions must be satisfied for a
SUSY Yang--Mills theory to be finite.  One of them is the
vanishing of the one--loop gauge $\beta$ function.  This
requirement constrains the spectrum of the theory essentially
fixing it (upto discrete possibilities), once the gauge group is
specified. A second requirement for finiteness is the vanishing of
all the anomalous mass dimensions of the chiral superfields at
one--loop. This would fix all the Yukawa couplings in terms of the
gauge coupling, at one--loop order.  This type of one--loop
finiteness also implies that the theory is finite to two loops
\cite{PW}. For the theory to be finite to all loops, the Yukawa
couplings must have unique power series expansions in terms of the
gauge coupling \cite{LPS}. With this condition satisfied, the
theory would have only one coupling -- the gauge coupling. The
Yukawa couplings are unified with the gauge couplings. This
``reduction of couplings'' is one of the key ingredients of
finiteness \cite{zim2,kubo1, kubo2, kubo3}. Certainly this makes
the idea of finiteness an attractive direction to pursue in
reducing the number of free parameters. One might hope that these
type of theories may arise from superstring theory.  Vanishing of
the $\beta$ functions lead to conformal invariance, which is one
of the cornerstones of string theory. Indeed there have been
several attempts to derive a grand unified theory from superstring
theory as its low energy 4-D limit (for example see \cite{witten,
anton}). The approach we adopt in this paper toward finiteness is
that of \cite{LPS}. Different approaches to finite theories can be
found, for example, in \cite{strassler, al}.

It will be extremely interesting to uncover finite theories that
are phenomenologically viable, at least in a broad sense. Attempts
have been made along this line with some success.  An immediate
question any finite theory should address is the consistency with
the observed masses and mixings of the quarks. The Yukawa
couplings are not arbitrary parameters in finite theories due to
the reduction of couplings.  The mass of the top--quark has been
predicted within finite theories, and shown to be in good
agreement with experiments \cite{Kubo}.  The masses of the lighter
generation quarks have also been consistently accommodated in this
context. However, the mixing between all three generations has not
been implemented successfully thus far.  This is the major point
we wish to address in the present paper.

We shall present three models based on finite SUSY $SU(5)$ theory
which can induce the correct pattern of quark mixing and masses.
Additional flavor symmetries are necessary to meet the criterion
for finiteness that the power series expansion of the Yukawa
couplings in terms of the gauge coupling be unique.  We find that
non--Abelian discrete symmetries are extremely useful here.
Abelian symmetries that we have tried were not sufficient to make
the expansion coefficients of the Yukawa couplings unique,
non--Abelian continuous symmetries such as $SU(2)$ and $SU(3)$ are
too restrictive to allow the needed Yukawa couplings.  One of the
examples that we present is based on permutation symmetry, the
other is based on the tetrahedral symmetry $A_4$.  We also present
a third example based on $S_4$ permutation symmetry in which the
Yukawa couplings have a one--parameter family of solutions in
terms of the gauge coupling.  We anticipate that this
arbitrariness may be removable by additional symmetries.  If not,
the proof of all--loop finiteness will not go through, although
the theory will be finite to two--loop order.  Such two--loop
finite theories have been studied in Ref. \cite{kaz}. It is not
clear if the finiteness of the theory will be maintained by higher
order corrections.  Although these  models with parametric
solutions for the Yukawa couplings are more flexible (and thus
less predictive) from the phenomenological point of view, we
consider the all--loop finite models to be more attractive.

In Section 2, we review briefly the conditions for finiteness,
starting from the renormalization group equations (RGE) for a
generic supersymmetric theory. From one of the criteria, namely
vanishing of the one loop gauge $\beta$ function, it is not hard
to see that finite models with phenomenologically favorable
particle spectrum can be found more easily in $SU(5)$ than in
other groups \cite{HPS}. Some general results of practical
interest are given for finite $SU(5)$ models. In Section 3 we
propose three models and analyze them in detail. In all cases we
show that realistic quark masses and mixing angles can be
generated.  This enables us to address more detailed questions
such as the decay rate of the proton, which is perhaps one of the
thorniest problems faced by SUSY GUTs.  Generically finite
theories are  problematic \cite{Desh}, we give some plausible
resolutions.  Our conclusions are given in Section 4.

\section{Finite Theories: A brief review}

The one loop gauge and Yukawa beta functions and the one loop
anomalous dimension of the matter fields in a generic SUSY
Yang--Mills theory are given by \cite{PW}:
\begin{eqnarray}\label{bg1}\beta^{(1)}_{g}&=&\frac{1}{16\pi^{2}}\left
(\sum_{R}
    T(R)-3C_{2}(G)\right)\\
\gamma^{(1)i}_{j}&=&\lambda^{ikl}\lambda_{jkl}-2C_{2}(R)g^{2}\delta^{i}_{j}
\\
\beta^{(1)}_{ijk}&=&\frac{1}{16\pi^{2}}[\lambda_{ijp}\gamma^{p}_{k}
+(k\leftrightarrow
i)+(k\leftrightarrow j)] \end{eqnarray}

\noindent where $T(R)$, $C_{2}(R)$ and $C_{2}(G)$ are the Dynkin
indices of the matter fields and the quadratic Casimirs of the
matter and gauge representations respectively. $\lambda^{ijk}$ and
$\beta^{(1)}_{ijk}$ are the Yukawa couplings and the one--loop
Yukawa $\beta$ function of $\lambda^{ijk}$. The criteria of all
loop finiteness for $N=1$ supersymmetric gauge theories can be
stated as follows \cite{LPS}: (i) It should be free from gauge
anomaly, (ii) the gauge $\beta$-function vanishes at one loop:
\begin{eqnarray}\label{bg2}\beta^{(1)}_{g}=0,\end{eqnarray}
(iii) there exists solution of the form $\lambda=\lambda(g)$ to
the conditions of vanishing one-loop anomalous  dimensions
\beq\label{gamma1}
 \gamma^{(1)i}_{j}=0,
\eeq and (iv) the solution is isolated and non-degenerate when
considered as a solution of vanishing one-loop Yukawa
$\beta$-function:
 \beq \beta^{(1)}_{ijk}=0.\hspace{2cm}\eeq
If all four conditions are satisfied, the dimensionless parameters
of the theory would depend on a single gauge coupling constant and
the $\beta$ functions will vanish to all orders.

The first step is to choose the gauge group. From (i), we see that
the vanishing of the one loop gauge $\beta$ function puts a strong
constraint on the particle content, leaving only discrete set of
possibilities. It seems hard to find phenomenologically viable
models other than in $SU(5)$ \cite{HPS}. For example, if one
chooses $SO(10)$ to be the gauge group, and tries to build a
finite model with necessary particle content in the traditional
sense, one quickly ``runs" out of the Dynkin indices: according to
Eq. (1) the sum of the Dynkin indices over the matter fields
should be equal to 24 in this case. On the other hand, field
content of the traditional $SO(10)$ GUT is $3\times {\bf 16}$ of
fermions, ${\bf 54}$, ${\bf 45}$, ${\bf 10}+{\bf 10^\prime}$ and
${\bf 16}+\overline{\bf 16}$ of Higgs, if the gauge symmetry is to
be broken by renormalizable terms in the superpotential. Then,
since the sum of the Dynkin indices of these fields is equal to
32, one ends up exceeding the gauge Dynkin index. Much the same
result can be reached for $SU(6)$ etc. While it will be of great
interest to uncover finite models other than $SU(5)$, here we will
confine ourselves to the case of finite $SU(5)$.

Beginning with the particle content of minimal SUSY $SU(5)$ theory
with three families of fermions belonging to  $3\times({\bf
10}+\bar{\bf 5})$, an adjoint {\bf 24} Higgs ($\Sigma$) and $({\bf
5}+\bar{\bf 5})$ Higgses one sees that vanishing of the one--loop
gauge $\beta$ function requires the introduction of additional
fields whose Dynkin indices add up to 3.  This happens if there
are three additional ${\bf 5}+\overline{\bf 5}$ matter fields,
which may be either Higgs--like bosonic fields or vector--like
fermionic fields.  This is in fact the most promising case from
phenomenology.  There are two other possibilities, viz.,  adding
${\bf 10} + \overline{\bf 10}$ or adding ${\bf 10} + {\bf 5} +
2\times \overline{\bf 5}$.  In the first case, realistic quark
masses cannot arise, in the second case one would be left with a
fourth family of fermions which remains light to the weak scale.
For phenomenological reasons we do not pursue these two
alternatives, and choose to work with $3 \times \{{\bf
5}+\overline{\bf 5}\}$ plus the minimal SUSY $SU(5)$ spectrum.

The finiteness criteria require that Eqs. (4)-(6) should give a
unique set of solutions to the Yukawa couplings. The equations are
linear in the square of the absolute values of the couplings.
Hence one expects the solutions to the Yukawa couplings to be
either zero or of order the gauge coupling. They are not free
parameters anymore. In order to satisfy the hierarchy in masses of
the fermions, one can choose the VEVs of the Higgs bosons to be
hierarchical. Naively this would need at least three Higgs
multiplets coupling to the up--quark sector, and three multiplets
coupling to the down--quark sector.  It is interesting that finite
$SU(5)$ spectrum admits the needed Higgs, which can be as many as
4 in each sector.  We will be interested in the case where at
least three of the ${\bf 5}+\overline{\bf 5}$ fields are
Higgs--like (viz., they develop VEVs of the order the electroweak
scale).  In fact, we shall see shortly that independent of this
phenomenoligical requirement, the vanishing of the one--loop
anomalous dimensions necessitates that at least three pairs of
${\bf 5}+\overline{\bf 5}$ have Yukawa couplings to the three
families of fermions. We will focus on inducing realistic mixing
among the three families of quarks, which has not been achieved in
earlier analyses \cite{Kubo}.

To search for a finite model, one has to write down a specific
superpotential and try to find a set of solutions satisfying the
criteria that all the Yukawa coupling wave function
renormalization factors vanish at one--loop.  We consider the
following superpotential (assuming an unbroken $R$--parity):

\begin{eqnarray}\label{sp1}
W&=&\sum^{3}_{i,j=1}\sum_{a}\left(\frac{1}{2}u^{a}_{ij}{\bf 10}_{i}{\bf 10}
_{j}H_{a}+d^{a}_{ij}{\bf 10}_{i}\bar{{\bf 5}}_{j}\bar{H}_{a}\right)
\nonumber\\
  &+&\sum_{ab}k^{ab}{H}_{a}\Sigma \bar{H}_{b}
  +\frac{\lambda}{3}\,{\Sigma}^{3}+f \,{\bf 5}\,\Sigma \,\bar{\bf 5}.
\end{eqnarray}
Here $i, j=(1-3)$ are family indices and $a$ and $b$ are Higgs
indices. $a$ and $b$ run from 1 to either 3 or 4. If it is up to
4, the last term is absent. $H$ and $\bar{H}$ denote the ${\bf 5}+
\overline{\bf 5}$ fields and $\Sigma$ the adjoint chiral matter
field responsible for the GUT symmetry breaking.  Note that in
order to have a successful doublet--triplet mass splitting, at
least one of the couplings $f$, $k^{ab}$ should be non--vanishing.

From Eq. (7), the anomalous dimensions of Eq. (2) can be written
in matrix form as:

\begin{eqnarray}
\label{g1}
\gamma_{{\bf 10}_{i}{\bf 10}_{j}}&=&3(u_{a}u^{\dag}_{a})_{ij}
+2(d_{a}d^{\dag}_{a})_{ij}-\frac{36}{5}g^{2}\delta_{ij}\nonumber\\
\gamma_{\bar{{\bf 5}}_{i}\bar{{\bf 5}}_{j}}&=&4(d^{\dag}_{a}d_{a})_{ij}
-\frac{24}{5}g^{2}\delta_{ij}\nonumber\\
\gamma_{H_{a}{H}_{b}}&=&3Tr(u^{\dag}_{a}u_{b})+\frac{24}{5}(kk^{\dag})_{ab}
-\frac{24}{5}g^{2}\delta_{ab}\nonumber\\
\gamma_{\bar{H}_{a}\bar{H}_{b}}&=&4Tr(d_{a}d^{\dag}_{b})
+\frac{24}{5}(k^{\dag}k)_{ab}-\frac{24}{5}g^{2}\delta_{ab}\\
\gamma_{{\bf 5}}&=&\gamma_{\bar{{\bf 5}}}=\frac{24}{5}f^{2}
-\frac{24}{5}g^{2}\nonumber\\
\gamma_{24}&=&Tr(k^{\dag}k)+f^{2}+\frac{21}{5}\lambda^{2}-10g^{2}.\nonumber
\end{eqnarray}

\noindent According to the third criteria, Eq. (\ref{gamma1}), in
order to have finite theory, all these anomalous mass dimension
have to be zero. Thus, the problem of finding a finite model
shifts to the problem of finding a set of solutions, where all the
anomalous dimensions in Eq. (\ref{g1}) vanish. Let us introduce a
new notations for the matrices:
\begin{eqnarray}\label{mt}
U\equiv u_{a}u^{\dag}_{a},&D\equiv d_{a}d^{\dag}_{a},\hspace{1cm}&
D^{\prime}\equiv d^{\dag}_{a}d_{a},\hspace{1cm}\tilde{U}_{ab}\equiv
Tr(u^{\dag}_{a}u_{b}),\nonumber\\
&\tilde{D}_{ab}\equiv Tr(d_{a}d^{\dag}_{b}),&\hspace{.5cm}K\equiv
k^{\dag}k,\hspace{1cm}\tilde{K}\equiv kk^{\dag},
\end{eqnarray}
where the trace is over the generation indices. From Eq. (8), it
follows that the number of $H$ fields coupling to ${\bf 10}_i{\bf
10}_j$ should be equal to the number of $\overline{H}$ fields
coupling to ${\bf 10}_i \overline{\bf 5}_j$ fields.  Furthermore,
at least 3 $H$ fields (and 3 $\overline{H}$ fields) must have such
couplings. To see this let us take the trace of the matrices of
the anomalous dimensions in Eq. (\ref{g1}) over both the fermionic
indices and the Higgs indices. One gets:
\begin{eqnarray}
3Tr(U)&+&2Tr(D)=3\times \frac{36}{5}g^{2}\nonumber\\
4Tr(D^{\prime})&=&3\times \frac{24}{5}g^{2}\nonumber\\
3Tr(\tilde{U})&+&\frac{24}{5}Tr(\tilde{K})=n_{H}\times
\frac{24}{5}g^{2}\\
4Tr(\tilde{D})&+&\frac{24}{5}Tr(K)=n_{\bar{H}}\times
\frac{24}{5}g^{2} ,\nonumber
\end{eqnarray}
where $n_H$ and $n_{\bar{H}}$ are the number of the Higgs fields
coupling to the three family of fermions in the up--sector and the
down--sector respectively. Subtracting the third equation from the
last in Eq. (10), we get
\begin{eqnarray}
4Tr(\tilde{D})-3Tr(\tilde{U})=(n_{\bar{H}}-n_{H})\times\frac{24}{5}g^{2}.
\nonumber
\end{eqnarray} Observing the following relation \begin{eqnarray}
Tr(U)=Tr(\tilde{U})\hspace{.3cm} \hspace{.3cm}
Tr(D)=Tr(D^{\prime})=Tr(\tilde{D}),\nonumber\end{eqnarray} one
finds that
\begin{eqnarray}\label{nb}n_{H}=n_{\bar{H}}.\end{eqnarray}
One can also see that the matrices $K$ and $\tilde{K}$ in Eq.
(\ref{mt}) vanish if $n_H = n_{\bar{H}}=3$. That is, $k_{ab} = 0$
for all $(a,b)$. Doublet--triplet splitting can be achieved in
this case since $f \neq 0$ is allowed.   If $n_H = n_{\bar{H}}
\leq 2$, no solution exists for Eq. (10).  We conclude that at
least three Higgs multiplets must couple to the fermion fields in
finite $SU(5)$.

Vanishing of the right--hand side of Eq. (8), needed for
finiteness, will in general lead to parametric solutions.  In
order to satisfy the condition for all--loop finiteness,
additional symmetries are usually necessary. Under these extra
symmetries different Higgs multiplets will have different charges,
which would prevent them from coupling to the same set of fermion
fields.  If two different Higgs multiplets $H_1$ and $H_2$ coupled
to the same fermion fields, say ${\bf 10}_1 {\bf 10}_2$, then
$\gamma_{H_1 H_2}$ will not vanish in general, and so the theory
will not be finite. We now present a classification of the Yukawa
coupling matrices which ensures in a simple way that the
off--diagonal entries of the anomalous dimension matrices are all
automatically zero.  While this classification is not the most
general, it can be applied to a wide class of models.  Let us
write the superpotential Eq. (\ref{sp1}) in the following form:
\begin{eqnarray}
W=\frac{1}{2}{\bf 10_i}{\bf 10_j}{V^u}_{ij}+{\bf 10_i}{\bf
\bar{5}_j}{V^d}_{ij}+.\,.\,.\, ,
\end{eqnarray}
where
\begin{eqnarray*}
{V^u}_{ij}=u^a_{ij}H_a,\,\,\,\,\, {V^d}_{ij}=d^a_{ij}\bar{H_a}~.
\end{eqnarray*}
The structures of $V^u$ matrices which automatically have all
off--diagonal anomalous dimensions to be zero is obtained as
follows.  Consider the case where three pairs of $(H,\bar{H})$
couple to the chiral families.  There are four distinct forms for
the matrix $V^u$:
\begin{eqnarray}V^{(1)}\equiv \pmatrix{u_{11}H_{1}&u_{12}H_3 &u_{13}H_2
\cr\nonumber
u_{12}H_3 &u_{22}H_2 &u_{23}H_1 \cr\nonumber
 u_{13}H_2 &u_{23}H_1 &u_{33}H_3 }\nonumber
\hspace{.5cm} V^{(2)}\equiv\pmatrix{u_{11}H_2 &u_{12}H_1
&0\cr\nonumber u_{12}H_1 &u_{22}H_3 &u_{23}H_2 \cr\nonumber
0&u_{23}H_2 &u_{33}H_3 }\nonumber\end{eqnarray} \vspace{.5cm}
\begin{eqnarray}\label{tex1}V^{(3)}\equiv\pmatrix{u_{11}H_3 &u_{12}H_1
&0\cr
u_{12}H_1 &u_{22}H_3 &u_{23}H_2 \cr 0&u_{23}H_2 &u_{33}H_3 }
\hspace{.5cm}V^{(4)}\equiv\pmatrix{u_{11}H_1 &0&0\cr 0&u_{22}H_3
&u_{23}H_2 \cr 0&u_{23}H_2 &u_{33}H_3 }~.
\end{eqnarray}
The form of $V^d$ in this case is identical to Eq. (13), except
that $u_{ij}$ is replaced by $d_{ij}$ and $H_i$ by $\bar{H}_i$.
While $V^u$ is a symmetric matrix, $V^d$ is asymmetric. Any given
Higgs field appears at most once in a given row or column in all
the matrices of Eq. (13). This guarantees that all off--diagonal
$\gamma$ function entries are zero. It can be shown that Eq. (13)
is the most general set of matrices that satisfy this constraint
(upto relabeling of generation number and Higgs number), provided
that there is no cancellation between various terms to generate a
zero in the off--diagonal $\gamma$ matrix.  It is possible that
such cancellations occur in the presence of non--Abelian flavor
symmetries, but not with Abelian symmetries.  Even for the case of
non--Abelian symmetries, we have found the classification of Eq.
(13) very useful.

If four pairs of $(H+\bar{H})$ couple to fermion families, the
matrix $V^u$ can have the following four structures (upto
relabeling of generation and Higgs indices):
\begin{eqnarray} V^{(1)}\equiv\pmatrix{u_{11}H_1 &u_{12}H_4 &u_{13}H_2
\cr\nonumber
u_{12}H_4 &u_{22}H_2 &u_{23}H_1 \cr\nonumber
 u_{13}H_2 &u_{23}H_1 &u_{33}H_3 }\nonumber
\hspace{.5cm}V^{(2)}\equiv\pmatrix{u_{11}H_2 &u_{12}H_1
&0\cr\nonumber u_{12}H_1 &u_{22}H_2 &u_{23}H_4 \cr\nonumber
0&u_{23}H_4 &u_{33}H_3 }\nonumber\end{eqnarray} \vspace{.5cm}
\begin{eqnarray}\label{tex2}V^{(3)}\equiv\pmatrix{u_{11}H_3 &u_{12}H_1
&u_{13}H_2 \cr
u_{12}H_1 &u_{22}H_2 &u_{23}H_4 \cr u_{13}H_2 &u_{23}H_4
&u_{33}H_3 } \hspace{.5cm}V^{(4)}\equiv\pmatrix{u_{11}H_3
&u_{12}H_1 &u_{13}H_2 \cr u_{12}H_1 &u_{22}H_3 &u_{23}H_4 \cr
u_{13}H_2 &u_{23}H_4 &u_{33}H_3 }~.
\end{eqnarray}
$V^d$ in this case will have similar structure, assuming that its
form is similar to $V^u$.  In all cases, one can easily verify
that the off--diagonal contributions to the anomalous dimension
matrices are all zero.

\section{The Quark Mixing in Finite GUT}

It is possible to find solutions for the vanishing of the
anomalous dimensions of Eq. (8) with the forms of $V^u$ and $V^d$
given as in Eq. (13)-(14). We have examined all possible cases,
including $V^u$ taking the form of $V^{(i)}$ while $V^d$ taking
the form of $V^{(j)}$ with $i$ and $j$ not necessarily the same.
We found parametric solutions wherein one or (typically) more
parameters are not determined.  That would forbid a unique
expansion of the Yukawa couplings in terms of the gauge coupling,
one of the requirements of finiteness.  It is possible to remove
this arbitrariness by imposing additional flavor symmetries. Three
examples of this type are proposed here and analyzed in detail.
In the first example, isolated non--degenerate solution to the
Yukawa couplings is obtained by imposing a $(Z_4)^3 \times P$
symmetry, where $P$ stands for permutation.  The second example is
based on the tetrahedral group $A_4$. A third example based on
$S_4$ symmetry is also presented, which actually gives a one
parameter family of solutions.  If this parameter is chosen to
have a specific value (which we believe can be enforced by a
symmetry) the solutions will again be isolated and
non--degenerate.

\subsection{$(Z_4)^3\times P$ Model}

Let us give the transformation properties of the fields under the
discrete symmetry we impose. The symmetries are $(Z_4)^3$,
identified as the $Z_4$ subgroup of generation number, and a
permutation symmetry acting on  both the fermion and the Higgs
generations. The fields transform under $(Z_4)^3$ as:

\begin{eqnarray}
 &&{\bf 10}_{1}:(i,1,1),\,\,\,\,\,\,\, {\bf 10}_{2}:(1,i,1),\,\,\,\,\,\,\,
{\bf 10}_{3}:(1,1,i),\nonumber\\
&&\bar{{\bf 5}}_{1}:(i,1,1),\,\,\,\,\,\,\,\,\,\,\bar{{\bf 5}}_{2}:(1,i,1),
\,\,\,\,\,\,\,\,\,\,\bar{{\bf 5}}_{3}:(1,1,i),\nonumber\\
&&(H_{1},\bar{H}_{1}):(-1,1,1),\,\,\,\,\,(H_{2}\bar{H}_{2}):(1,-i,-i),\\
&&(H_{3},\bar{H}_{3}):(1,1,-1),\,\,\,\,\,(H_{4},\bar{H}_{4}):(1,-i,-i).
\nonumber
\end{eqnarray}

\noindent The action of the permutation symmetry $P$ on the fields
is as follows:

\begin{center}
\begin{tabular}{llll}
${\bf 10}_{1}\leftrightarrow {\bf 10}_{3}$,& $\bar{{\bf
5}}_{1}\leftrightarrow\bar{{\bf 5}}_{3}$,&$H_{1}\leftrightarrow
H_{3}$,&$ \bar{H}_{1}\leftrightarrow \bar{H}_{3}$,\\
${\bf 10}_{2}\leftrightarrow {\bf 10}_{2}$,&$\bar{{\bf
5}}_{2}\leftrightarrow\bar{{\bf 5}}_{2}$,&$H_{2}\leftrightarrow
H_{4}$,& $\bar{H}_{2}\leftrightarrow \bar{H}_{4}.$
\end{tabular}\end{center}

\noindent The most general $SU(5)\times (Z_4)^3 \times P$ \,
invariant superpotential is:
\begin{eqnarray}
W&=&a ({\bf 10}_1 {\bf 10}_1 H_1+{\bf 10}_3 {\bf 10}_3 H_3)+ b
({\bf 10}_1 {\bf 10}_2 H_4+{\bf 10}_2 {\bf 10}_3 H_2)\nonumber\\
&+& c ({\bf 10}_1 \bar{{\bf 5}}_1 \bar{H}_1+{\bf 10}_3 \bar{{\bf
5}}_3 \bar{H}_3)
+d({\bf 10}_1 \bar{{\bf 5}}_2 \bar{H}_4+{\bf 10}_3 \bar{{\bf 5}}_2 \bar{H}
_2)\\
&+&e({\bf 10}_2 \bar{{\bf 5}}_1 \bar{H}_4+{\bf 10}_2 \bar{{\bf
5}}_3 \bar{H}_2) + k(H_1 \bar{H}_1 \Sigma + H_3 \bar{H}_3
\Sigma)+\frac{\lambda}{3}\Sigma^3.\nonumber
\end{eqnarray}
The matrices $V^u$ and $V^d$ (defined in Eq. (\ref{tex2})) for
this model are then:
\begin{eqnarray}
V^u=\pmatrix{a\,H_1&b\,H_4&0 \cr b\,H_4&0&b\,H_2\cr
0&b\,H_2&a\,H_3} \hspace{1cm}
V^d=\pmatrix{c\,\bar{H}_1&d\,\bar{H}_4&0\cr e\, \bar{H}_4&0&e\,
\bar{H}_2\cr 0&d\,\bar{H_2}&c\,\bar{H}_3},
\end{eqnarray} and the coupling matrix of the Higgs fields to the adjoint
field is given by:
\begin{eqnarray*}K=diag(k, 0, k, 0).\end{eqnarray*}

Note that all superpotential couplings can be made real by field
redefinitions. One can then take all parameters of Eq. (16) to be
real and positive.  This is an important point for the solution to
be non--degenerate.

The condition (iii) of the criteria for finiteness (vanishing of
all the anomalous dimensions) leads to the following simple system
of equations:
\begin{eqnarray}
3(a^2+b^2)+2(c^2+d^2)=\frac{36}{5}g^2,&&3(2b^2)+2(2e^2)=\frac{36}{5}g^2,
\nonumber\\
c^2+e^2=\frac{6}{5}g^2,&&2b^2=\frac{8}{5}g^2,\\
a^2+\frac{8}{5}k^2=\frac{8}{5}g^2,&&d^2+e^2=\frac{6}{5}g^2,\nonumber\\
2d^2=\frac{6}{5}g^2,&&a^2+\frac{6}{5}k^2=\frac{6}{5}g^2.\nonumber
\end{eqnarray} This gives a unique solution which is isolated and
non-degenerate:
 \beq\label{solz4} \left(a^2,
b^2, c^2, d^2, e^2, k^2, \lambda^2\right)=\left(\frac{4}{5}g^2,
\frac{4}{5}g^2, \frac{3}{5}g^2, \frac{3}{5}g^2,\frac{3}{5}g^2,
\frac{1}{2}g^2, \frac{15}{7}g^2\right). \eeq There is no sign
ambiguity for the Yukawa couplings themselves, since they have all
been made real and positive.

Let us now turn to the question of comparing the predictions of
Eq. (19) with experiments.  First of all, all three families of
quarks mix with one another, so realistic CKM mixings become
possible, unlike earlier attempts within finite GUTs.  Setting the
overall factor $a\langle H_3\rangle = 1$, we can write the mass
matrix $M^u$ for the up--type quarks in the following form:
\begin{eqnarray}\label{mz4}
M^u = \pmatrix{c^u_{11}\epsilon_u^4&c^u_{12}\epsilon_u^3&0\cr
c^u_{21}\epsilon_u^3&0&\epsilon_u\cr
0&\epsilon_u&1},\end{eqnarray} where
$\epsilon_u\equiv\frac{\langle H_2\rangle}{\langle H_3\rangle}$,
$c^u_{11}\epsilon_u^4\equiv\frac{\langle H_1\rangle}{\langle
H_3\rangle}$ and $c^u_{12}\epsilon_u^3\equiv\frac{\langle
H_4\rangle}{\langle H_3\rangle}$.  The mass matrix for the
down--type quarks, $M^d$, has a similar form, with $\epsilon_u$
replaced by $\epsilon_d$ and $c_{ij}^u$ replaced by $c_{ij}^d$.
These matrices are generalizations of the Fritszch form. Note that
Eq. (20) is a special case of texture $V^{(2)}$ in Eq.
(\ref{tex2}). The mass eigenvalues are obtained in the
approximation $\epsilon_u \ll 1,~c_{ij}^u \sim 1$ to be:
\begin{eqnarray}\label{eig1}
&&m_u\simeq\left(c^u_{11}+\left(c^u_{12}\right)^2\right)
\epsilon_u^4\nonumber\\
&&m_c\simeq-\epsilon_u^2+\left(1-c^u_{12}\right)^2\epsilon_u^4\nonumber\\
&&m_{t}\simeq1+\epsilon_u^2-\epsilon_u^4
\end{eqnarray}
in units of $a\left\langle H_3 \right\rangle$.  Similar
expressions hold in the down--type quark sector.  The CKM mixing
elements are then given by:
\begin{eqnarray}\label{fit2}
&&V_{us}=c^u_{12}\epsilon_u-c^d_{12}\epsilon_d+O(\epsilon^3)\nonumber\\
&&V_{cb}=\epsilon_d-\epsilon_u+O(\epsilon^3)\\
&&V_{ub}=c^u_{12}\epsilon_u\epsilon_d-c^u_{12}\epsilon^2_u+O(\epsilon^4),
\nonumber
\end{eqnarray}
where $\epsilon_d$ and $c^d_{12}$ correspond to the down quark
sector.   (For simplicity, we have assumed all parameters to be
real.  This assumption is not necessary, realistic CP violation
can also arise from Eq. (20).)

Observe that the mass hierarchy between generations can be
accommodated in this model by assuming a hierarchy in the VEVs of
the Higgs doublets.  We have in mind a scenario where only one
pair of Higgs doublets survive below the GUT scale, to be
identified as $H_u$ and $H_d$ of MSSM.  These are linear
combinations of all four of the original Higgs doublets.  That
would enable all $H_i$ ($i=1-4$) to acquire VEVs.  The $H_u$ field
of MSSM is dominantly $H_3$, but has small (of order $\epsilon_u$)
component of $H_2$ in it, and even smaller components of $H_4$ (of
order $\epsilon_u^3$) and $H_1$ (of order $\epsilon_u^4$) in it.
These amounts are dictated by the bilinear terms in the
superpotential involving $H_i$ and $\bar{H}_i$ fields ($W \sim
m_{ij} H_i \bar{H}_j$).  These bilinear terms are assumed to break
the $(Z_4)^3 \times P$ symmetry softly.  We see that the desired
mass hierarchy is reproduced in this way.

Since the Yukawa couplings of the third generation quarks are
fixed in this model in terms of the gauge coupling, the mass of
the top quark and the parameter $\tan\beta$ are determined.  Let
us denote the MSSM Yukawa couplings of the top and the bottom
quarks to $H_u$ and $H_d$ fields to be $y_t$ and $y_b$
respectively.  To a good approximation, $H_u$ is $H_3$ and $H_d$
is $\bar{H}_3$. Thus we see from Eq. (19) that $y_t \simeq
(\sqrt{4/5})g$ and $y_b \simeq (\sqrt{3/5})g$, both of which are
fixed in terms of $\alpha_G \simeq 1/25$.  We now extrapolate
these Yukawa couplings to the weak scale using the MSSM
renormalization group equations. The top quark mass and the
parameter $\tan\beta$ can be predicted using the relations
\begin{eqnarray}
&&m_t=m_b\frac{y_t}{y_b}\sqrt{{y_b}^2\frac{v^2}{{m_b}^2}-1} \simeq y_t v
\nonumber\\
 &&\tan\beta=\frac{m_t}{m_b}\frac{y_b}{y_t},
\end{eqnarray}
where $v=174$ GeV. With $m_b(m_b)$ taken to be 4.4 GeV and with
$\alpha_3(m_Z) = 0.118$ we find the numerical values to be:
 \begin{eqnarray}&&m_t=174~GeV\nonumber\\
&&\tan\beta=53.\end{eqnarray} The predicted value of $m_t$ is
nicely consistent with the experimentally determined value,
$\tan\beta$ tends to be large in this class of models.

There is one other non--trivial prediction in this model, because
of the zeros present in Eq. (20).  We take it to be a prediction
for the strange quark mass.  From Eqs. (21) and (22) we find
$m_s(1$ GeV$) \simeq 80$ MeV, if we take $V_{cb} \simeq 0.043,
m_b(m_b) = 4.4$ GeV, $m_c(m_c) = 1.37$ GeV, $V_{us} = 0.22$ and
$V_{ub} = 0.004$.  This value of $m_s$ is on the low side, but may
be consistent with recent lattice evaluations \cite{brah}.  We
also note that since $\tan\beta$ is predicted to be large, the
finite threshold corrections to $V_{cb}$ through chargino--stop
exchange is significant \cite{raby}.  This could modify $V_{cb}$
by as much as 30\%.  With a 30\% reduction in $V_{cb}$ arising
from this diagram, we predict $m_s(1~{\rm GeV}) \simeq 100$ MeV,
which is quite acceptable.

%The numerical values of the mixing angles are found to be
%\begin{eqnarray}
%V_{us}&=&\nonumber\\
%V_{cb}&=&\\
%V_{ub}&=&\nonumber\end{eqnarray} The value of $V_{cb}$ being
%bigger than the experimentally measured value \cite{PDG}, is a
%common feature of this texture.
% Here we have $M_{GUT}=$

\subsection{$A_4$ Model}

Now we present a second model that leads to realistic quark
mixings and masses.  It is based on $SU(5)\times A_4$ symmetry.
$A_4$ is the group of even permutations of four objects. It is the
symmetry group of a regular tetrahedron.  This group has
irreducible representations (denoted by the dimensions) $1$,
$1^\prime$, $1^{\prime\prime}$ and $3$.  The $1^\prime$ and
$1^{\prime \prime}$ are complex conjugate of each other. The
product $3 \times 3$ decomposes as
\begin{eqnarray}
3\times 3=1+1^\prime+1^{\prime\prime}+3+3.
\end{eqnarray}
If we denote the components of 3 as (a,b,c), the various terms are
given by \cite{ma}:
\begin{eqnarray}\label{alg1}
1&=&a_1 a_2+b_1 b_2 +c_1 c_2\nonumber\\
1^\prime&=&a_1 a_2+\omega^2b_1 b_2 +\omega c_1 c_2\\
1^{\prime \prime}&=&a_1 a_2+\omega b_1 b_2 +\omega^2c_1
c_2,\nonumber
\end{eqnarray}
where $\omega=e^{2i\pi/3}$. (Note that $1+\omega+\omega^2=0$.)

The transformations properties of the fields of $SU(5)$ under
$A_4$ are:
\begin{eqnarray}
{\bf 10}_{i}&:&3\hspace{.5cm}\hspace{1cm}{\bf
\bar{5}}_{i}\hspace{.2cm}:\hspace{.3cm}3\\\nonumber
(H_{a},H_{4})&:&3+1^{\prime}\hspace{.6cm}(
\bar{H}_{a},\bar{H}_{4})\hspace{.3cm}:\hspace{.3cm}3+1^{\prime}\hspace
{.6cm}
\Sigma=1,
\end{eqnarray}
where $i=1\div3$ and $a=1\div3$. Using the algebra presented in
Eq. (\ref{alg1}), the superpotential invariant under $SU(5)\times
A_4$ symmetry is:
\begin{eqnarray}W&=&a({\bf 10}_1 {\bf 10}_1 +\omega {\bf 10}_2 {\bf 10}_2
+\omega^2 {\bf 10}_3
{\bf 10}_3)H_4\nonumber\\
&+&c({\bf 10}_1 \bar{{\bf 5}}_1 +\omega {\bf 10}_2 \bar{{\bf 5}}_2
+\omega^2 \bar{{\bf 5}}_3 {\bf 10}_3)\bar{H}_4\nonumber\\
&+&b[({\bf 10}_1 {\bf 10}_2 +{\bf 10}_2 {\bf 10}_1)H_1+({\bf 10}_1 {\bf 10}
_3 +{\bf 10}_3 {\bf 10}_1 )H_2+({\bf 10}_2 {\bf 10}_3+{\bf 10}_3 {\bf 10}
_2)H_3]\nonumber\\
&+&d[({\bf 10}_1 \bar{{\bf 5}}_2 +{\bf 10}_2 \bar{{\bf 5}}_1)\bar{H}_1+d
({\bf 10}_1 \bar{{\bf 5}}_3 +{\bf 10}_3 \bar{{\bf 5}}_1 )\bar{H}_2+({\bf
10}_2 \bar{{\bf 5}}_3+{\bf 10}_3 \bar{{\bf 5}}_2)\bar{H}_3]\nonumber\\
&+&k(\bar{H}_{1}H_{1}+\bar{H}_{2}H_{2}+\bar{H}_{3}H_{3})\Sigma+\frac
{\lambda}{3}\Sigma^3.\nonumber
\end{eqnarray}
By field redefinition the $\omega$ factors can be removed from
$W$. Actually, all the coupling constants in Eq. (28) can be made
real and positive. The condition of vanishing anomalous dimensions
for this model can be written as follows:
\begin{eqnarray}
&&3(a^2+2b^2)+2(c^2+2d^2)=\frac{36}{5}g^2\nonumber\\
&&4(c^2+2d^2)=\frac{24}{5}g^2\nonumber\\
&&3(3a^2)=\frac{24}{5}g^2\\
&&3(2b^2)+\frac{24}{5}k^2=\frac{24}{5}g^2\nonumber\\
&&4(3c^2)=\frac{24}{5}g^2\nonumber\\
&&4(2d^2)+\frac{24}{5}k^2=\frac{24}{5}g^2.\nonumber\end{eqnarray}
This gives the following isolated and non-degenerate solution:
\begin{eqnarray} a^2=b^2=\frac{8}{15}g^2,\hspace{2mm}
c^2=d^2=\frac{2}{5}g^2,\hspace{2mm}k^2=\frac{1}{3}g^2.\end{eqnarray}
The resulting up--quark mass matrix can be written as:
\begin{eqnarray}
M^u = \sqrt{\frac{8}{15}}g\langle
H_4\rangle\pmatrix{1&1+\epsilon_1&1+\epsilon_2\cr
1+\epsilon_1&1&1+\epsilon_3\cr1+\epsilon_2&1+\epsilon_3&1},
\end{eqnarray}
where
\begin{eqnarray}
\epsilon_1=\frac{\langle H_1\rangle}{\langle
H_4\rangle}-1,\,\,\,\,\epsilon_2=\frac{\langle H_2\rangle}{\langle
H_4\rangle}-1,\,\,\,\,\epsilon_3=\frac{\langle H_3\rangle}{\langle
H_4\rangle}-1,
\end{eqnarray}
with a similar form for the down--type quark matrix. We can
accommodate the mass hierarchy by taking $\epsilon_{1,2,3} \ll1$.
This structure has been considered in \cite{Fish}, where it has
been shown to agree well with experimental data.

In the $A_4$ model, since all $H_i$ have almost equal VEVs,
$\left\langle H_4 \right \rangle \simeq \left\langle
H_u\right\rangle/2$.  Furthermore, from Eq. (30), we have $m_t
\simeq 3\sqrt{8/15}g\left\langle H_4 \right\rangle$, so that $y_t
= (\sqrt{6/5})g$ at the GUT scale.  Similarly, $y_b =
(\sqrt{9/10})g$ at the GUT scale. These boundary conditions lead
to the predictions

\begin{eqnarray}m_t&=& 177~{\rm GeV}\nonumber\\
\tan\beta&=& 53.\end{eqnarray} As shown in Ref. \cite{Fish}, all
the quark mixing angles can be correctly reproduced in this model.

\subsection{$S_{4}$ Model}

We now present a third example based on $S_4$ symmetry. This
symmetry alone would lead to a one parameter family of solutions
for the Yukawa couplings.  Although we have not found a symmetry
that will uniquely fix this parameter, we suspect that such a
symmetry might actually exist.  Keeping this in mind, we proceed
to analyze this model.  $S_4$ is the permutation symmetry
operating on four objects. It has the following irreducible
representations: $(1,~1',~2,~3,~3')$ \cite{Pakvasa}. We choose the
following assignment of the chiral superfields under $S_4$:
\begin{eqnarray}
{\bf
10}_{i}&:&3,\hspace{1cm}(H_{a},H_{4})\hspace{.2cm}:\hspace{.2cm}3+1,\hspace
{.6cm}\hspace{.6cm}
\Sigma:1\nonumber,\\
\bar{{\bf
5}}_{i}&:&3,\hspace{1cm}(\bar{H}_{a},\bar{H}_{4})\hspace{.2cm}:\hspace
{.2cm}3+1,
\end{eqnarray}
The superpotential invariant under this symmetry is
\begin{eqnarray}
W&=&a[({\bf 10}_1 {\bf 10}_3 +{\bf 10}_3 {\bf 10}_1)H_1 +({\bf
10}_2 {\bf 10}_3 +{\bf 10}_3 {\bf 10}_2)H_2 +({\bf 10}_1 {\bf
10}_1 -{\bf 10}_2 {\bf 10}_2)H_1]\nonumber\\&+&b({\bf 10}_1 {\bf
10}_1 +{\bf 10}_2 {\bf 10}_2 +{\bf 10}_3 {\bf
10}_3)H_4\nonumber\\&+&c[({\bf 10}_1 \bar{{\bf 5}}_3 +{\bf 10}_3
\bar{{\bf 5}}_1)\bar{H}_1+({\bf 10}_2 \bar{{\bf 5}}_3 +{\bf 10}_3
\bar{{\bf 5}}_2)\bar{H}_2+({\bf 10}_1 \bar{{\bf 5}}_1 -{\bf 10}_2
\bar{{\bf 5}}_2)\bar{H}_4]\nonumber\\&+&d({\bf 10}_1 \bar{{\bf
5}}_1 +{\bf 10}_2 \bar{{\bf 5}}_2+{\bf 10}_3 \bar{{\bf
5}}_3)\bar{H}_4\\&+&k (H_1 \bar{H}_1+H_2 \bar{H}_2+H_3
\bar{H}_3)\Sigma \nonumber\\&+&k_4 H_4 \bar{H}_4 \Sigma +
\frac{\lambda}{3}\Sigma^3.\nonumber
\end{eqnarray}
The $V^u$ and $V^d$ matrices that arise from this superpotential
are:
\begin{eqnarray}
V^u&&\pmatrix{a\langle H_3\rangle +b \langle
H_{4}\rangle&0&a\langle H_1 \rangle\cr 0&b \langle H_4\rangle&a
\langle H_2\rangle\cr a \langle H_1\rangle&a\langle
H_2\rangle&-a\langle H_3\rangle +b\langle H_4\rangle},
\hspace{1cm}\end{eqnarray}
\begin{eqnarray}&&V^d =\pmatrix{c \langle \bar{H}_3\rangle+d\langle\bar{ H}
_4\rangle&0&c
\langle \bar{H}_1\rangle\cr 0&d\langle \bar{H}_4\rangle&c \langle
\bar{H}_2\rangle\cr c \langle \bar{H}_1\rangle&c
\langle\bar{H}_2\rangle&-c \langle \bar{H}_3\rangle+ d \langle
\bar{H}_3\rangle}.\end{eqnarray} The coupling matrix $k$
connecting the Higgs fields ($H,\bar{H}$) and the adjoint field
$\Sigma$ is:
\begin{eqnarray*}k=diag(k, k, k, k_4).\end{eqnarray*}
The condition for vanishing anomalous mass dimensions is then:

\begin{eqnarray} &&3(2a^2+b^2)+2(2c^2+d^2)=\frac{36}{5}g^2\nonumber\\
&&4(2c^2+2d^2)=\frac{24}{5}g^2 \nonumber\\
&&4(2d^2)+\frac{24}{5}k^2=\frac{24}{5}g^2\nonumber\\
&&4(3d^2)+\frac{24}{5}{k_4}^2=\frac{24}{5}g^2\\
&&3(2a^2)+\frac{24}{5}k^2=\frac{24}{5}g^2\nonumber\\
&&3(2b^2)+\frac{24}{5}{k_4}^2=\frac{24}{5}g^2\nonumber\\
&&3k^2+{k_4}^2+\frac{21}{5}\lambda^2=\frac{24}{5}g^2.
\nonumber\end{eqnarray}
The solution to this set of equations has one free parameter. We
choose it be $k_4$, in which case the solution is:
\begin{eqnarray} \left(a^2, b^2, c^2,
d^2\right)&=&\left(\frac{8}{15}g^2+\frac{4}{15}{k_4}^2,\,\,
\frac{8}{15}g^2-\frac{4}{15}{k_4}^2,\,\,
\frac{2}{15}g^2+\frac{1}{5}{k_4}^2,\,\,
\frac{2}{15}g^2-\frac{1}{5}{k_4}^2\right)\nonumber\\
\left( e^2, k^2,
\lambda^2\right)&=&\left(\frac{2}{5}g^2-\frac{2}{5}{k_4}^2,\,\,
\frac{1}{3}g^2-\frac{1}{3}{k_4}^2,\,\,
\frac{15}{7}g^2\right).\end{eqnarray} To eliminate this
undetermined parameter $k_4$ one needs to introduce an additional
symmetry. A $Z_2$ symmetry can set $k_4=0$, but if this $Z_2$
commutes with $S_4$, it will also set some other parameters to be
zero.  We suspect a $Z_2$ that does not commute with the $S_4$
symmetry might set $k_4$ equal to zero, while preserving the
solution Eq. (39).  We find the model phenomenologically
interesting for this case. The mass matrix is for the up--type
quarks is:
\begin{eqnarray}
M^u = \sqrt{\frac{8}{15}}\,\,g\,\pmatrix{\langle H_3\rangle +
\langle H_{4}\rangle&0&\langle H_1\rangle \cr 0& \langle
H_4\rangle& \langle H_2\rangle\cr \langle H_1\rangle&\langle
H_2\rangle&-\langle H_3\rangle +\langle H_4\rangle},\end{eqnarray}
and a similar form for the down--type quarks.

To explain the mass hierarchy, we first set the $(1, 1)$ entry of
the mass matrices both in the up and the down sectors to be zero
by choosing $\langle H_3\rangle$ and $\langle H_4\rangle$ as:
\begin{eqnarray}\langle H_3\rangle+\langle H_4\rangle\sim 0,
&&\langle \bar{H}_3\rangle+\langle \bar{H}_4\rangle\sim
0.\nonumber\end{eqnarray} Furthermore, we take $\left\langle
H_1\right\rangle$ and  $\left\langle \bar{H}_1\right\rangle$ to be
smaller than $\left\langle H_2\right\rangle \sim \left\langle
H_4\right\rangle$. One immediate observation from the structure is
that the rotation between the second and the third generations is
large. These large rotations from the up and the down sectors will
cancel out. Let us define $\frac{\langle H_2\rangle}{\langle
H_4\rangle}=\sqrt{2}(1+\delta^u)$. In the limit
$\epsilon^u\equiv\frac{\langle H_1\rangle}{\langle
H_4\rangle}\rightarrow 0$, the rotation in the second and third
generations is:
\begin{eqnarray}\pmatrix{1&\sqrt{2}(1+\delta^u)\cr\sqrt{2}(1+\delta^u)&2}.
\end{eqnarray}
Form this one finds
\begin{eqnarray}&&\frac{m_c}{m_t}=-\frac{4}{9}\delta^u,\end{eqnarray}
where $m_c$ and $m_t$ are the masses of charm and top quarks. The
rotation  angle is:
\begin{eqnarray}
\tan(2\theta^u_{23})=2\sqrt{2}(1+\delta^u).\end{eqnarray} The
large rotation angle will cancel out in $V_{cb}$, leaving only the
smaller corrections proportional to $\delta^{u,d}$. The large
rotation in the 2-3 space will induce an entry equal to
$\epsilon^u\sin\theta_{23}^u \left\langle H_4\right\rangle$ in the
(1,3) element.  From this, we obtain the following relations for
the quark mixing angles:
\begin{eqnarray}
&&V_{cb}=\frac{1}{2\sqrt{2}}\left|\frac{m_s}{m_b}\pm\frac{m_c}{m_t}
\right|\nonumber\\
&&V_{us}=\left|\sqrt{\frac{m_d}{ m_s}}\pm\sqrt{\frac{m_u}{ m_c}}\right|\\
&&V_{ub}=2\sqrt{2}\left|\frac{\sqrt{m_d
m_s}}{m_b}\pm\frac{\sqrt{m_u
m_c}}{m_t}\right|.\nonumber\end{eqnarray} These are of the right
order of magnitude, although in detail, the magnitude of $V_{cb}$
is somewhat smaller than what is needed and $V_{ub}$ is on the
larger side.  We consider this general agreement with experiments
to be encouraging.

\section{Concluding remarks}

We have presented in this paper several models for quark masses
and mixings in the context of finite $SU(5)$ GUT.  These theories
are attractive candidates for an underlying theory, since the
$\beta$ functions for the gauge and Yukawa couplings vanish to all
orders in perturbation theory.  The requirements on the theory to
be finite also leads to Yukawa--gauge unification, leading to a
single coupling constant in the theory.

The models presented are based on non--Abelian discrete symmetry,
which seem to be necessary to obtain isolated and non--degenerate
solutions to the Yukaw couplings when expressed as power series in
terms of the gauge coupling.  We find it interesting that
realistic quark masses and mixings can be generated in such a
framework.

There are several open questions, many of which cannot be
addressed until after finding a consistent quark mixing scheme. An
important question finite theories should address is how to avoid
rapid proton decay.  Because all the Yukawa couplings, including
those of the light generations,
 are order of $g$, color triplet exchange will generate a large amplitude
for proton decay through $d=5$ operators \cite{Desh}. This may
simply be a technical problem associated with using $SU(5)$ as the
gauge group. One can envision other groups without the color
triplets, although no realistic model of this type are known to
us.  Within finite $SU(5)$, there are ways to suppress the
troublesome proton decay operators.  For example, if the SUSY
particle spectrum is such that the gauginos are light (of order
100 GeV), while the squarks are very heavy (of order 100 TeV or
larger), the $d=5$ proton decay problem goes away.  Although this
choice may not be that attractive from the point of view of
solving the gauge hierarchy problem, we emphasize that finiteness
of the theory says nothing per se about the scale of SUSY
breaking. A third alternative is to suppose that the masses of all
the extra color triplets in the theory are much heavier than the
GUT scale, even larger than the Planck scale.

In the framework of $SU(5)$ finite GUT, the following question
arises naturally: Is it possible to generate small neutrino
masses? If right--handed neutrinos are introduced as $SU(5)$
singlets, they can have no Yukawa couplings with the other fields,
due to the demand of finiteness.  We mention two possibilities to
induce small neutrino masses.  One is through  bilinear $R$-
parity violating terms of order the weak scale \cite{rpnu}. That
does not contradict the requirements of finiteness.  Another
possibility is to make use of Planck suppressed higher dimensional
operators, which can be constructed within finite $SU(5)$.

As we have shown in Sec. 2, within finite $SU(5)$, all four pairs
of ${\bf 5}+\overline{\bf 5}$ fields present in the theory must be
Higgs--like.  This is needed for achieving doublet--triplet
splitting.  If one pair were fermionic, the bad mass relations of
$SU(5)$, viz., $m_s = m_\mu$ and $m_d=m_e$ could have been
corrected by terms such as $\bar{\bf 5}_i {\bf 5}$ bilinear mass
terms along with $\bar{\bf 5} \Sigma {\bf 5}$ coupling.  Since
that is not possible, one has rely on either Planck suppressed
operators or finite gaugino diagrams to split the masses of
leptons versus down type quarks \cite{murayama}.  Both
possibilities appear to be viable.

\section*{Acknowledgments}

This work is supported in part by DOE Grant \# DE-FG03-98ER-41076,
a grant from the Research Corporation and by DOE Grant \#
DE-FG02-01ER-45684.


\begin{thebibliography}{99}

\bibitem{PW} A.J. Parkes and P.C. West,  Phys. Lett.{\bf B138} (1984) 99;
Nucl. Phys. {\bf B256} (1985) 340; P. West,
Phys. Lett. {\bf B137} (1984) 371; D.R.T. Jones and A.J. Parkes,
Phys. Lett. {\bf B160} (1985) 267; D.R.T. Jones and L. Mezinescu,
Phys. Lett. {\bf B136} (1984) 242; {\bf B138} (1984) 293; A.J.
Parkes, Phys. Lett. {\bf B156} (1985) 73.

\bibitem{LPS} C. Lucchesi, O. Piguet and K. Sibold, Helv. Phys. Acta {\bf
61} (1988) 321;
 Phys. Lett. {\bf B201} (1988) 241: C. Lucchesi and G. Zoupanos,
Fortsch.Phys.{\bf 45} (1997)129.

\bibitem{zim2} W. Zimmermann,  Com. Math. Phys. {\bf 97} (1985) 211;
 R. Oehme and W. Zimmermann  Com. Math. Phys. {\bf 97} (1985) 569; R.
Oehme, K. Sibold and W. Zimmermann,
 Phys. Lett. {\bf B147} (1984) 117; {\bf B153} (1985) 142; R. Oehme,  Prog.
Theor. Phys. Suppl. {\bf 86} (1986)
 215.

\bibitem{kubo1} J. Kubo, K. Sibold and W. Zimmermann, Nucl. Phys. {\bf
B259} (1985) 331;
 Phys. Lett. {\bf B200} (1989) 185; K.S. Babu and S. Nandi,  Preprint
OSU-RN-202, UR-1056,
ER-13065-533, May 1988 (unpublished).

\bibitem{kubo2} J. Kubo,  Phys. Lett. {\bf B262} (1991) 472.

\bibitem{kubo3} J. Kubo, K. Sibold and W. Zimmermann, Phys. Lett. {\bf
B200} (1989) 185.

\bibitem{witten}E. Witten, Nucl. Phys. {\bf B268} (1986) 79; B. R. Greene,
K. H. Kirklin and
P. J. Miron, Nucl. Phys. {\bf B274} (1986) 574.

\bibitem{anton}I. Antoniadis and G.K. Leontaris, Phys. Lett. {\bf B216}
(1989) 333; G. Leontaris and N. Tracas, Z.
Phys. {\bf C56} (1992) 479;  Phys. Lett. {\bf B291} (1992) 44.

\bibitem{strassler} R.G. Leigh and M.J. Strassler,  Nucl. Phys. {\bf B447}
(1995) 95.

\bibitem{al}A.V. Ermushev, D.I. Kazakov and O.V. Tarasov, Nucl. Phys. {\bf
B281} (1987) 72; D.I. Kazakov,
 Mod. Phys. Let. {\bf A2} (1987) 663; Phys. Lett. {\bf B179} (1986)
 352.


\bibitem{Kubo} J. Kubo, M. Mondrag\'on and G. Zoupanos,  Nucl. Phys. {\bf
B424} (1994) 291;
 J.. Kubo, M. Mondrag{\' o}n, N.D. Tracas and G.Zoupanos,  Phys. Lett. {\bf
B342} (1995) 155; Kapetanakis, M.
Mondrag{\' o}n and G. Zoupanos,  Zeit. f. Phys. {\bf C60} (1993)
181; M. Mondrag{\' o}n and G. Zoupanos,  Nucl. Phys. {\bf B}(Proc.
Suppl.) {\bf 37C} (1995) 98; J. Kubo, M. Mondragon and G.
Zoupanos, Acta Phys.Polon.{\bf B27} (1997) 3911;


\bibitem{kaz}D.I. Kazakov and I.N. Kondrashuk, Int. J. Mod. Phys. {\bf A7}
(1992) 3869;
D.I. Kazakov, M.Yu. Kalmykov, I.N. Kondrashuk and A.V. Gladyshev,
Nucl.Phys. {\bf B471} (1996) 389.

\bibitem{HPS}
S. Hamidi, J. Patera and J.H. Schwarz, Phys. Lett. {\bf B141}
(1984) 349; X.D. Jiang and X.J. Zhou,
 Phys. Lett. {\bf B197} (1987) 156; {\bf B216} (1985) 160.

\bibitem{Desh} N. Deshpande, X-G. He and E. Keith,  Phys. Lett. {\bf B332}
(1994) 88.

\bibitem{brah}See for e.g. B. Brahmachari, Phys. Lett. {\bf B486} (2000)
118.


\bibitem{raby}T. Blazek, S. Raby and S. Pokorski, Phys. Rev. {\bf D52}
(1994) 4151; K.S. Babu and C. Kolda, Phys. Rev. Lett. {\bf 84}
(2000) 228.


\bibitem{ma} E. Ma and G. Rajasekaran, Phys. Rev. {\bf D64} (2001) 113012.

\bibitem{Fish} P. M. Fishbane and P. Kaus, Phys. Rev. {\bf D49} (1994)
4780; P.M. Fishbane and P. Kaus, Z.Phys.{\bf C75} (1997) 1.


\bibitem{Pakvasa} Y. Yamanaka, H. Sugawara and S. Pakvasa,  Phys. Rev. D
{\bf25}(1982) 1895.

\bibitem{rpnu}
See for instance: C.S. Aulakh, R.N. Mohapatra, Phys. Lett. {\bf
B119}, 136 (1982); L.J. Hall, M. Suzuki, Nucl. Phys. {\bf B231},
419 (1984).

\bibitem{murayama} J. Diaz-Cruz, H. Murayama and A. Pierce,
Phys. Rev. {\bf D65} (2002) 075011.


\end{thebibliography}
\end{document}